\documentclass[slac_one]{revtex4}
\usepackage{graphicx}
\usepackage{fancyhdr}
\begin{document}

{\hbox to\hsize {\hfill UFIFT-HEP-05-10 }}

\title{{\small{2005 International Linear Collider Workshop - Stanford,
U.S.A.}}\\ 
\vspace{12pt}
Slepton Mass Measurements at the LHC} 

%

\author{A. Birkedal\footnote{This talk was given by A. Birkedal, describing ongoing work performed in collaboration with the other authors.}, R.C.Group and K. Matchev}
\affiliation{University of Florida, Gainesville, FL 32611, USA}

\begin{abstract}
We first discuss the motivation for measuring slepton masses at the LHC, emphasizing their importance for cosmology.
Next we investigate the possibility of making slepton mass determinations at the LHC in neutralino decays.  
We demonstrate that by studying the shape of the dilepton invariant mass distribution in the decay
$\tilde\chi^0_2\to\tilde\chi^0_1 \ell^+\ell^-$, one can determine whether the slepton is real or virtual.
Furthermore, in case of virtual sleptons, it is possible to bound the slepton mass within a limited range.
We conclude by discussing future avenues for investigation.
\end{abstract}

\maketitle

\thispagestyle{fancy}


\section{INTRODUCTION} 
Low-energy supersymmetry remains the best motivated extension of the standard model.
The search for superpartners is among the primary objectives of the LHC experiments.
At the LHC, strong production of colored superpartners (squarks and the gluino) dominates
and there is an extensive body of literature devoted to the associated signatures.
In contrast, the direct production of non-colored superpartners (e.g. sleptons)
is less abundant, posing a challenge for their discovery~\cite{Sleptons, Andreev:2004qq}.  A recent analysis~\cite{Andreev:2004qq} finds that with $30\, {\rm fb}^{-1}$ of data CMS can discover right-handed sleptons up to $200$ GeV and left-handed sleptons up to $300$ GeV.

Supersymmetric theories preserving R-parity also generically contain viable WIMP (Weakly Interacting Massive Particle) 
dark matter candidates -- typically the lightest neutralino $\tilde\chi^0_1$. 
The discovery signatures in that case contain missing transverse energy due to
two stable $\tilde\chi^0_1$s in each event escaping the detector. The observation of a missing 
energy signal at the LHC will fuel the WIMP hypothesis.  However, a missing energy signal at a collider
only implies that particles have been created that are stable on a timescale that is characteristic of the detector size.
In order to prove that the missing energy particle is indeed viable dark matter, one needs to calculate its relic abundance.  To this end, one needs to measure all parameters which enter this calculation.  

The relic abundance of a dark matter particle is determined in large part by its annihilation cross section $\sigma\left(\chi \chi \rightarrow \sum_{i} X_{i}\right)$.  We have used $\chi$ to represent a generic dark matter particle.  $X_{i}$ is any allowed final state.  The post-WMAP determination of the dark matter abundance is accurate to about $10\%$~\cite{WMAP}.  Assuming a standard cosmology, one can then deduce a value for the cross section $\sigma\left(\chi \chi \rightarrow \sum_{i} X_{i}\right)$.   This, in turn, can be translated into a model-independent prediction for the rates of $e^+ e^- \rightarrow \chi \chi \gamma$, $q \bar{q} \rightarrow \chi \chi \gamma$, and $q \bar{q} \rightarrow \chi \chi \tilde{g}$ at colliders~\cite{ModInd}.  However, these searches are challenging at both the ILC~\cite{ModInd} and LHC~\cite{SWIMPDetect}.

In typical models, the slepton masses are among the key parameters 
in determining whether $\tilde\chi^0_1$ is a good dark matter candidate~\cite{AnnSigVExpr}.  For example, if the slepton is light, 
then slepton mediated annihilation diagrams are important.  In this case knowledge of the slepton mass is required 
to determine the relic abundance.  Conversely, if the slepton is heavy, its mass is unimportant for the relic abundance 
calculation~\cite{StringDerived,NonUniversal,FocusPoint}.  But without a collider measurement of the slepton mass, there may be a significant uncertainty 
in the relic abundance calculation.  This uncertainty results because the mass should then be allowed to vary within 
the whole experimentally allowed range.

To summarize, the importance of slepton discovery is two-fold.  First, supersymmetry predicts a superpartner for every standard model particle.  Therefore, the discovery of the superpartners of the leptons is an important step in verifying supersymmetry.  Second, knowledge of slepton masses is {\it always} important for an accurate determination of the relic abundance of $\tilde\chi_1^0$.  


In this talk we show that the LHC will indeed have sensitivity to slepton masses, even in the case of heavy sleptons.  We describe in detail how slepton masses can be determined from neutralino decays.  In the example of minimal supergravity (mSUGRA), we illustrate how this analysis is done in practice.  We show that the difference between real and virtual sleptons can be clearly seen.  We emphasize that establishing the presence of a real slepton in a cascade decay by our method is equivalent to a slepton discovery.  We also show that in case of virtual sleptons, one can frequently limit the allowed range of their masses, which is equivalent to a rough indirect slepton mass measurement.

\section{SLEPTON PHENOMENOLOGY}

\subsection{Sleptons at the LHC and ILC}
Direct production of sleptons at the LHC and LC are very similar.  For all sleptons except for selectrons at ILC, direct production proceeds through Drell-Yan processes.  However, the ability of the two colliders to measure slepton properties are very different.  At the linear collider, the fixed center of mass energy allows for accurate measurements of slepton masses up to the kinematical limit of the collider.  Lepton endpoints from slepton decays can be used to accurately determine slepton and neutralino masses.  Furthermore, threshold scans provide a very accurate alternative method for slepton mass measurement.  Beam polarization allows for a determination of the handedness of the sleptons ($\tilde{\ell}_L$ or $\tilde{\ell}_R$).  Finally, the background to these signatures can be easily removed with appropriate cuts.

In contrast, slepton studies at the LHC are challenging.  Most importantly, direct slepton production suffers from large backgrounds, mostly due to $W^+ W^-$ and $t \bar{t}$ production~\cite{Andreev:2004qq}.  The methods for slepton mass determination used at the linear collider are not applicable here.  Therefore, one must be more clever.  Fortunately, sleptons are produced in sizable quantities at the LHC through cascade decays.  These events can be easily triggered on and separated from the standard model backgrounds.  So in principle, these slepton events present an opportunity for a slepton mass measurement since sleptons can appear in the decays of gauginos.  A common situation in supersymmetric models is the hierarchy $|M_1 | < |M_2 | < |\mu |$.  In that case, sleptons affect the decay $\tilde\chi_2^0 \rightarrow \ell^\pm \ell^\mp \tilde\chi_1^0$.  The resulting dilepton distribution, in principle, contains information about the slepton mass $m_{\tilde{\ell}}$.  This situation is complicated by the fact that $\tilde\chi_2^0$ can also decay through a real or virtual $Z$: $\tilde\chi_2^0 \rightarrow Z \tilde\chi_1^0 \rightarrow \ell^\pm \ell^\mp \tilde\chi_1^0$.  In the next subsection we will investigate the process $\tilde\chi_2^0 \rightarrow \ell^\pm \ell^\mp \tilde\chi_1^0$ in detail.

\subsection{Slepton Masses through Neutralino Decays}
What is the observable in these events that is sensitive to the slepton mass?  In this analysis we will consider the dilepton invariant mass distribution, $m_{\ell \ell}$.  It is already well-known that the endpoint of $m_{\ell \ell}$ contains information about the masses of the real particles involved in the decay~\cite{Dilep}.  
\begin{itemize}
\item If the decay occurs through a real $Z$, $\tilde\chi_2^0 \rightarrow Z \tilde\chi_1^0 \rightarrow \ell^\pm \ell^\mp \tilde\chi_1^0$, then almost all such events will occur in the $Z$ mass peak, and endpoint information will be lost.
\item In the case of a virtual intermediate particle ($\tilde\chi_2^0 \rightarrow Z^* \tilde\chi_1^0 \rightarrow \ell^\pm \ell^\mp \tilde\chi_1^0$ or $\tilde\chi_2^0 \rightarrow \tilde{\ell}^{\pm *} \ell^{\mp} \rightarrow \ell^\pm \ell^\mp \tilde\chi_1^0$), this process is a three-body decay and the endpoint value is:

\begin{equation}
m_{\ell\ell,max} = m_{\tilde\chi_{2}^{0}} - m_{\tilde\chi_{1}^{0}}.
\label{virtual}
\end{equation}
\item Finally, if the decay is through a real slepton ($\tilde\chi_2^0 \rightarrow \ell^\pm \tilde{\ell}^\mp \rightarrow \ell^\pm \ell^\mp \tilde\chi_1^0$), the endpoint is at:

\begin{equation}
m_{\ell\ell,max} = \sqrt{\frac{\left(m_{\tilde\chi_{2}^{0}}^2-m_{\tilde{\ell}}^2\right)\left(m_{\tilde{\ell}}^2 - m_{\tilde\chi_{1}^{0}}^2\right)}{m_{\tilde{\ell}}^2}}.
\label{real}
\end{equation}
\end{itemize}

\begin{figure*}[t]
\centering
\includegraphics[width=100mm]{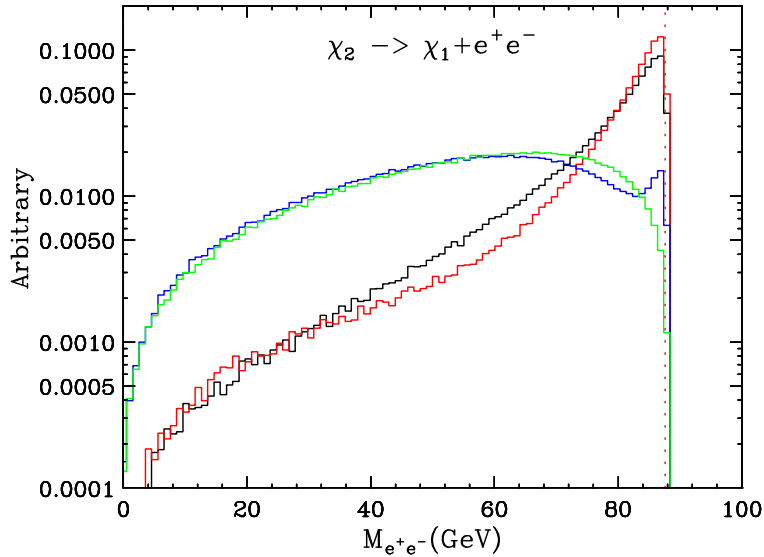}
\caption{Comparison of $M_{e^+ e^-}$ distributions for different selectron masses.  We only consider the diagrams mediated by $Z$ and $\tilde{e}_R$.  All parameters are held fixed except for the $\tilde{e}_R$ mass.  The (green, blue, red, black) line is for a ($300$, $500$, $1000$ GeV, and $\infty$) mass selectron, respectively.  The neutralino masses are kept constant, and their difference is $88$ GeV.} \label{InvMassDists}
\end{figure*}

The endpoint can be measured; however, its interpretation is ambiguous since we don't know {\it a priori} which formula is applicable (Eqn.~\ref{virtual} or Eqn.~\ref{real}).  But there is more information contained in the $m_{\ell\ell}$ distribution than just in the endpoint.  One would expect the shapes of the $Z$ and $\tilde{\ell}$ mediated distributions to be different.  Furthermore, the shape of the total decay distribution (including both $Z$ and $\tilde{\ell}$ contributions) should change as a function of the slepton mass.  Indeed this is the case, as illustrated in Figure~\ref{InvMassDists}.  In this Figure we show the dilepton invariant mass distribution resulting from the interference of the $Z$ and $\tilde{e}_R$-mediated diagrams.  Since the kinematic endpoint is kept fixed, this illustrates that endpoint analyses are largely insensitive to the slepton mass.  In all four cases, the slepton is virtual, but we see a clear difference in the shape of the distribution.  This gives us hope that virtual slepton masses can be determined by studying the shape of the decay distributions.  In the case of two body decay through a real slepton, the $m_{\ell\ell}$ distribution will be triangular, as required by phase space.

\section{SLEPTON MASS MEASUREMENTS AT THE LHC}
\subsection{The Kolmogorov-Smirnov Test}
Figure~\ref{InvMassDists} makes it clear that the $m_{\ell\ell}$ distribution changes significantly as a function of slepton mass.  How does one use this fact to statistically differentiate between distributions coming from sleptons of different mass?  Or, how does one statistically discriminate between an experimentally measured distribution and template distributions generated with a variety of slepton masses?  The Kolmogorov-Smirnov~\cite{KolSmir} test provides a statistical procedure for differentiating between such distributions.

The Kolmogorov-Smirnov test calculates the maximum cumulative deviation between two unit-normalized distributions.  It then translates this number into a confidence level with which it can be excluded that the two distributions come from the same underlying distribution.  So, the Kolmogorov-Smirnov test cannot tell that two distributions came from the same underlying distribution, but it {\it can} tell that two distributions {\it did not} come from the same underlying distribution.  We use this test in the next section to explicitly investigate the ability of the LHC to determine slepton masses in mSUGRA.

\subsection{Slepton Masses in mSUGRA}

We now demonstrate a slepton mass measurement at the LHC in mSUGRA parameter space.  For this demonstration, we select events that contain the decay of $\tilde\chi_{2}^{0} \rightarrow e^+ e^- \tilde\chi_{1}^{0}$.  The standard model background is efficiently suppressed by a combination of the standard cuts, including $M_{eff}$, jet $p_T$, missing energy and lepton $p_T$~\cite{Dilep}.  Additionally, opposite flavor distribution subtraction should eliminate most SUSY background~\cite{OppSub}.  However, we emphasize that these reasonable assumptions are in the process of being confirmed.  All relevant background will be included in the forthcoming preprint~\cite{UsPreprint}.

\begin{figure*}[t]
\centering
\includegraphics[width=85mm]{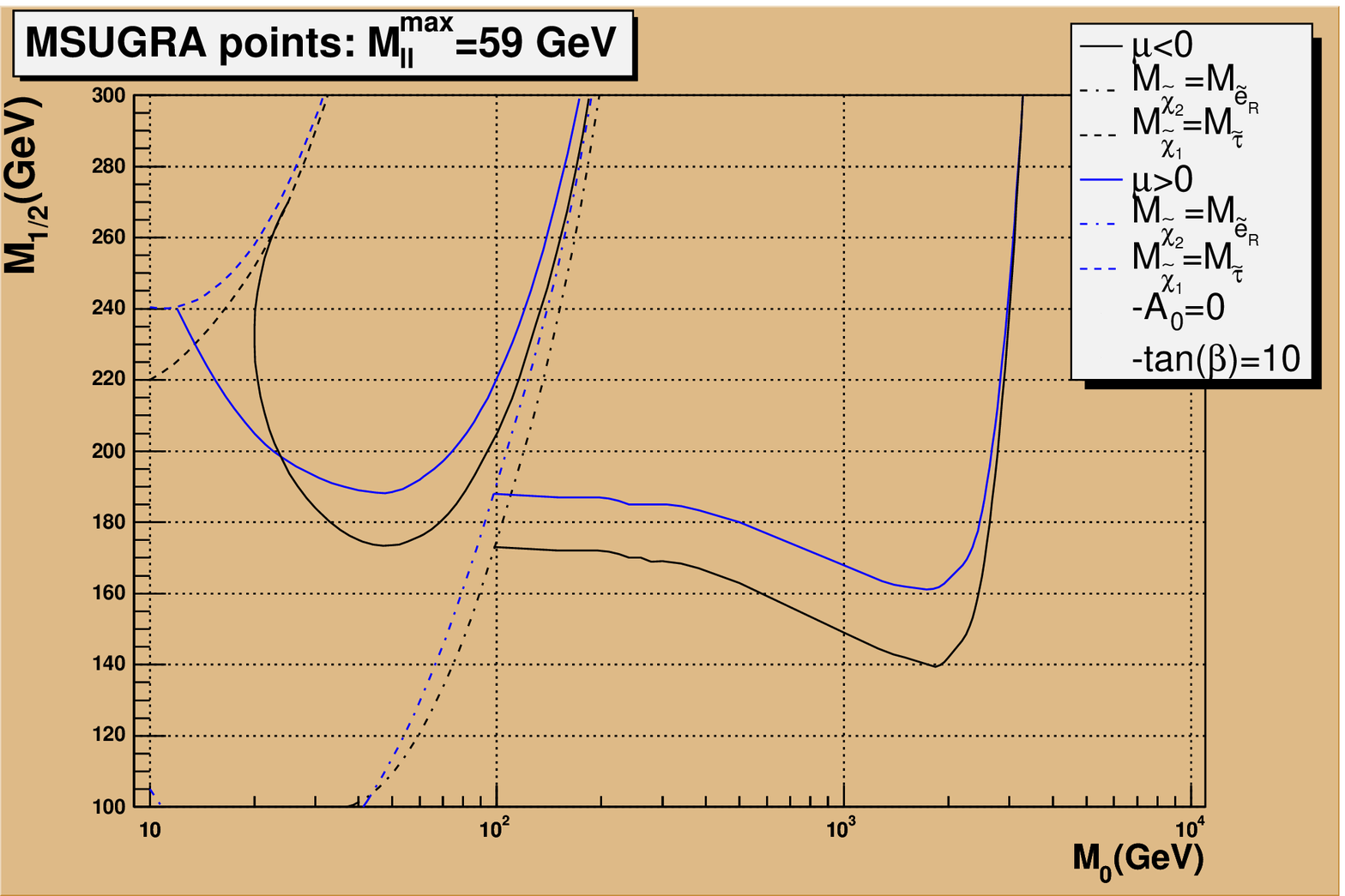}
\includegraphics[width=85mm]{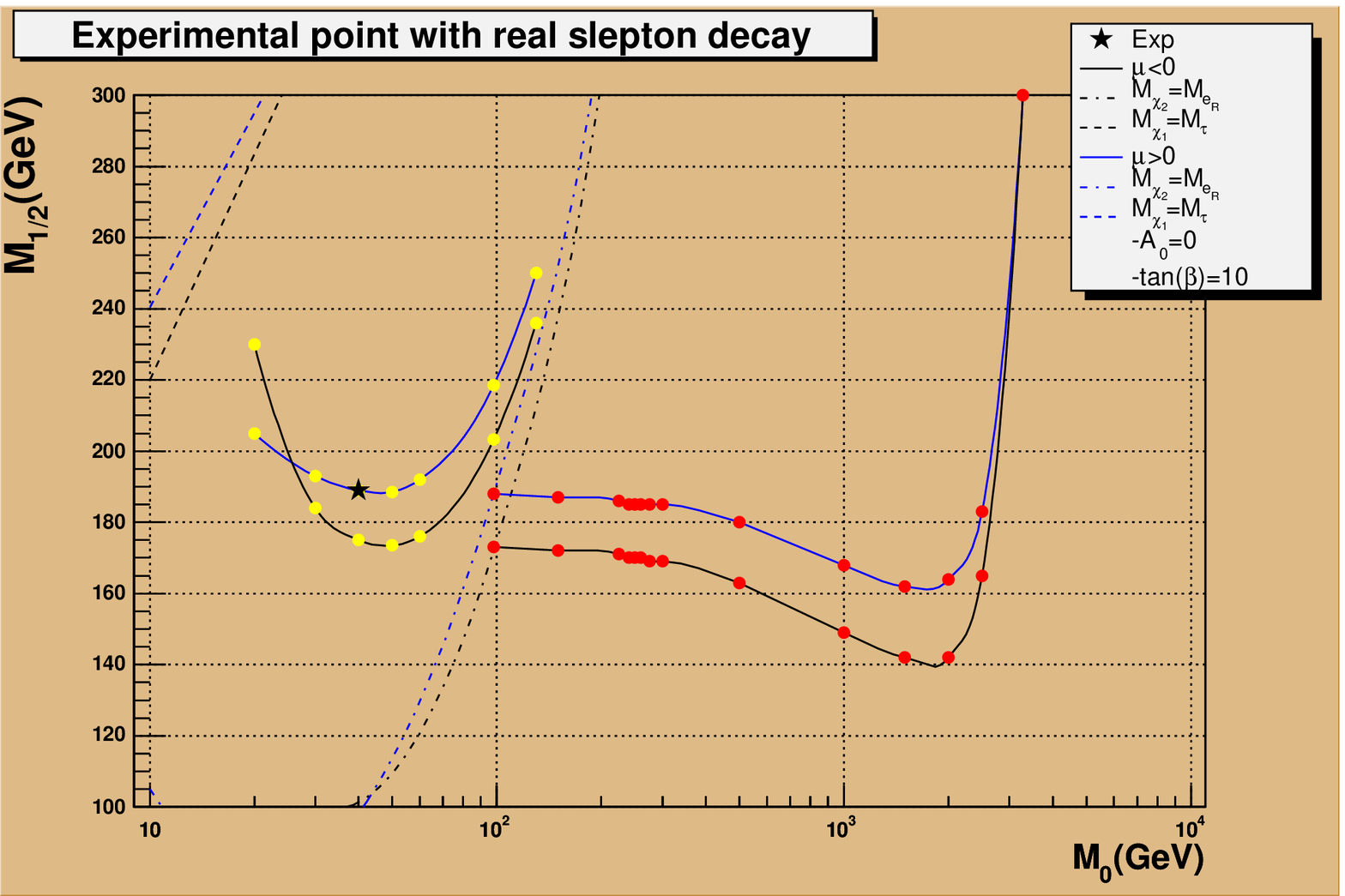}
\includegraphics[width=85mm]{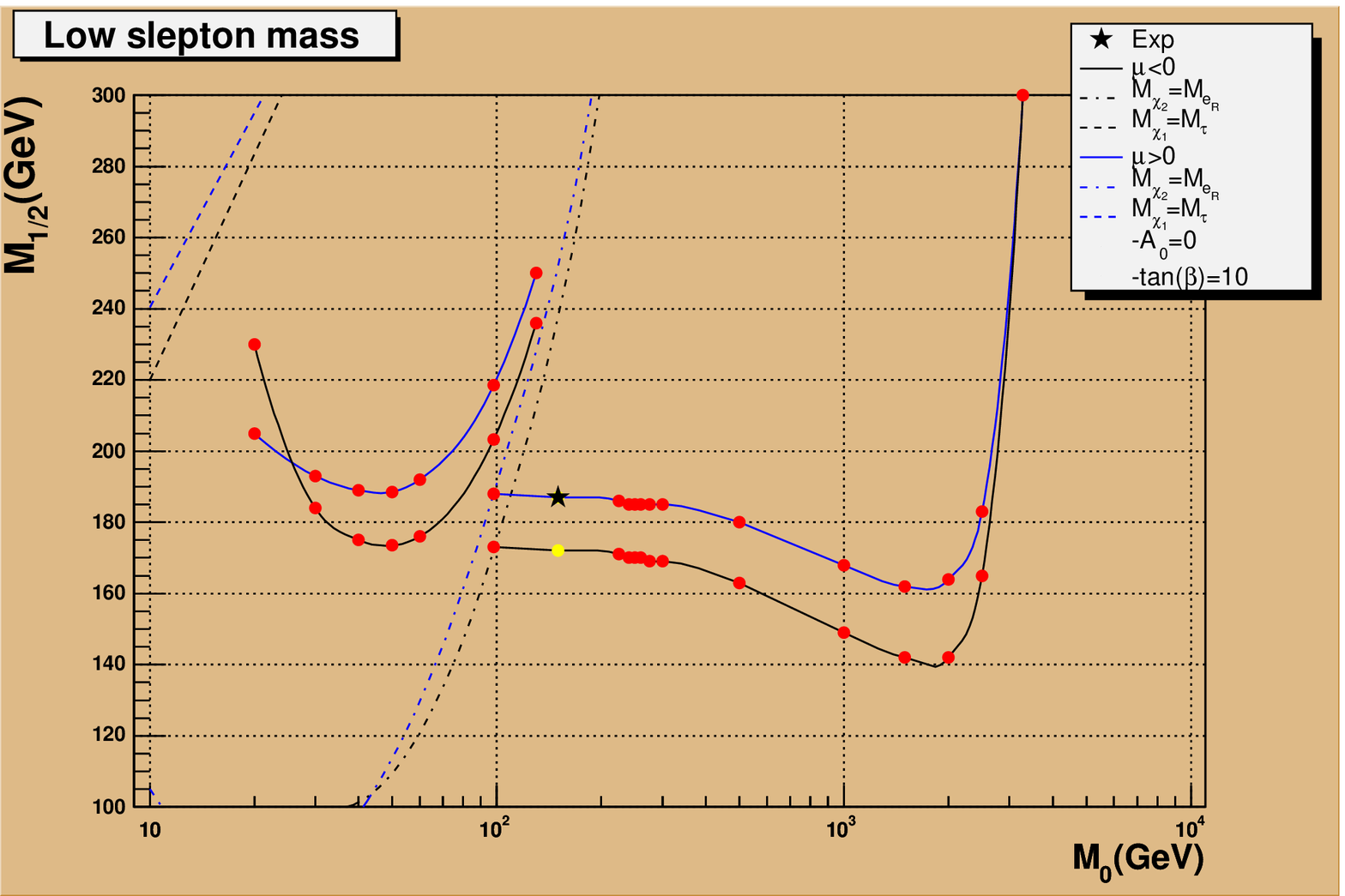}
\includegraphics[width=85mm]{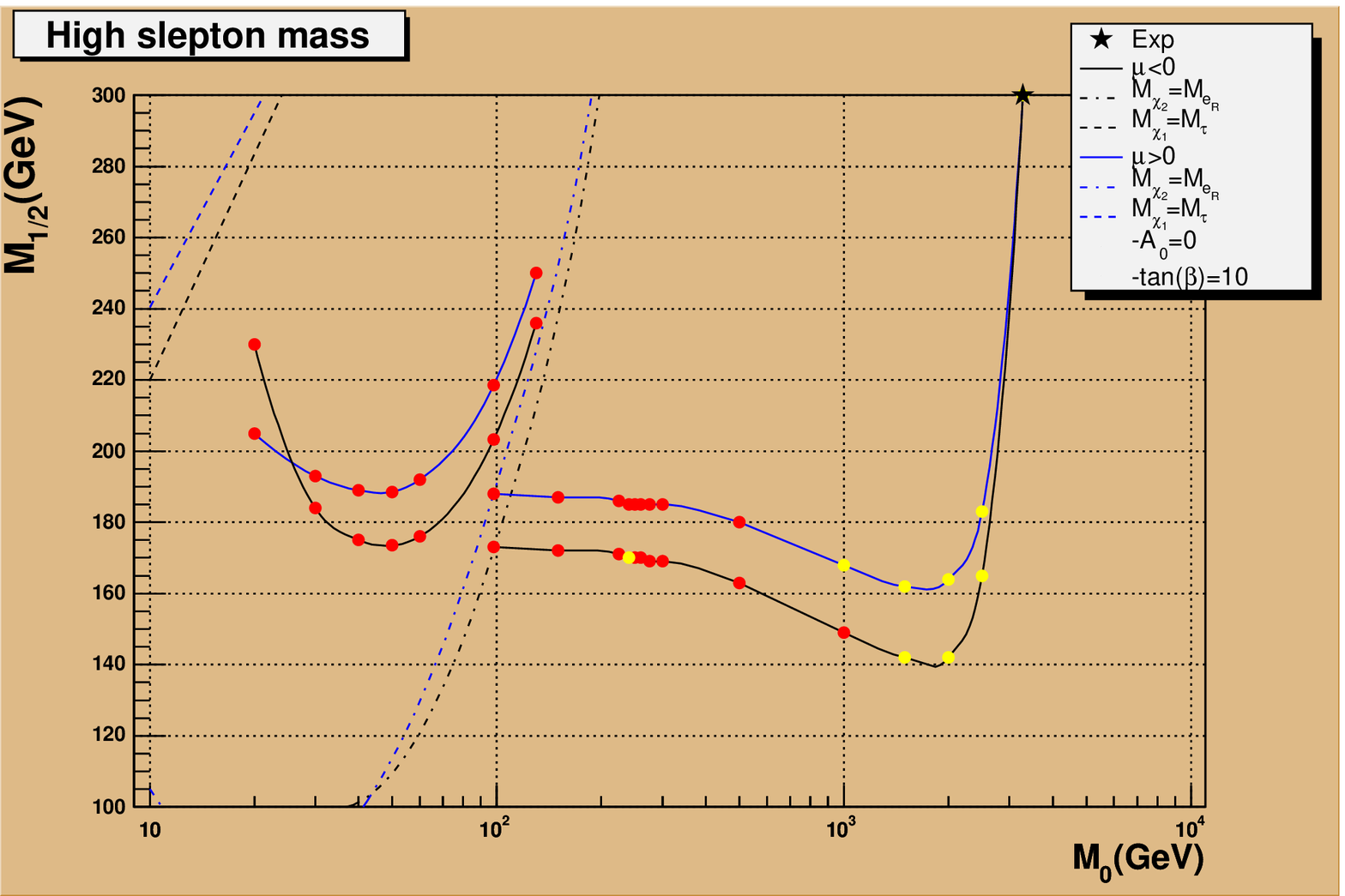}
\caption{Slepton mass determination in a slice of mSUGRA parameter space with $A_0 = 0$ and $\tan\beta=10$.  Here $M_0$ ($M_{1/2}$) is the universal scalar (gaugino) mass parameter.  The upper left plot shows the effect on the mSUGRA parameter space of fixing the dilepton kinematic endpoint of the $\chi_{2}^{0} \rightarrow e^+ e^- \chi_{1}^{0}$ decay to be $m_{\ell \ell,max} = 59$ GeV.  The other three plots show the ability of the Kolmogorov-Smirnov test to bound the mass of the lightest selectron.  Details are explained in the text.} \label{DistPlots}
\end{figure*}

First, we imagine that the LHC experiments have observed the dilepton mass distribution and have measured a kinematic endpoint at $59$ GeV.  What are the implications of this measurement for the SUSY mass spectrum?  Generally speaking, this reduces the parameter space by one degree of freedom.  This is illustrated in the upper left plot in Figure~\ref{DistPlots}.  In this plot, we show a two dimensional slice of the mSUGRA parameter space by fixing $A_0 = 0$ and $\tan\beta=10$.  The measurement of the kinematic endpoint reduces the two dimensional parameter space to one dimensional line segments.  In the Figure, these are the solid lines.  In mSUGRA, there is also the binary choice of $\mu >0$ or $\mu<0$, and we show the corresponding results in blue and black, respectively.  The dashed lines in the upper left corner indicate where $m_{\tilde\chi_{1}^{0}} = m_{\tilde{\tau}_1}$.  Any points to the left of these lines are ruled out by constraints on charged dark matter.  The dashed lines running through the middle of the plot indicate where $m_{\tilde{e}_R} = m_{\tilde\chi_{2}^{0}}$.  This is where the slepton-mediated neutralino decays changes from being three body ($\tilde{e}_R$ is virtual to the right of these lines) to two body ($\tilde{e}_R$ is real to the left of these lines).

In the other three plots in Figure~\ref{DistPlots}, we display the power of the Kolmogorov-Smirnov test in identifying the mass range of the slepton.  In each plot, we have taken one point (denoted by the black star) as resulting from a possible experimental measurement.  We have assumed $1000$ data events in the dilepton distribution.  Then we have compared this 'experimental data' with template distributions (denoted by the red and yellow dots).  The red dots indicate points that the Kolmogorov-Smirnov test can identify as not coming from the same distribution as our 'experimental data' at the $95\%$ confidence level.  The yellow dots denote points that could not be excluded at the $95\%$ confidence level.

The upper right plot of Figure~\ref{DistPlots} takes a point with a real slepton as the 'experimental data.'  Here we can see that the Kolmogorov-Smirnov test can easily distinguish between a real slepton and a virtual slepton.  This is an important distinction.  First, it is a slepton discovery.  Second, it allows us to put a solid bound on the slepton mass and, thirdly,  also unambiguously determines which endpoint equation (Eqn.~\ref{virtual} or Eqn.~\ref{real}) should be used.  The lower left plot takes a point with a low-mass virtual slepton as the 'experimental data.'  For this point, the Kolmogorov-Smirnov test rules out at the $95\%$ confidence level all of the example template points except for the one with the same value of $M_0$, but with the opposite sign of $\mu$.  Thus, the mass of the slepton can be determined with great accuracy even though it is not produced as a real particle.  Finally, the 'experimental data' in the lower right plot is taken from the focus point region~\cite{FocusPoint} with extremely heavy sleptons.  For this point, all of the sleptons are around $3$ TeV.  For $\mu > 0$, the Kolmogorov-Smirnov test is able to put a lower bound on the slepton mass of around $1$ TeV, as can be seen from the left-most yellow point on the blue line.  For $\mu < 0$, the generic lower bound is even tighter, around $1.5$ TeV, except for one point near $250$ GeV.  In this region for $\mu < 0$ there is significant cancellation between the slepton-mediated and $Z$-mediated decay amplitudes, so the Kolmogorov-Smirnov test is unable to distinguish this point from the point with heavy sleptons.
\section{CONCLUSION}
In this talk, we have motivated the importance of measuring slepton masses at the LHC.  An inability to significantly bound the slepton masses introduces large uncertainties in any subsequent calculation of the neutralino dark matter abundance.  We have also identified dilepton invariant mass distributions from neturalino decays as one avenue for determining slepton masses at the LHC.  The Kolmogorov-Smirnov test provides a statistical method of discriminating between dilepton invariant mass distributions resulting from different mass sleptons.  We investigated the decay $\tilde\chi_{2}^{0} \rightarrow e^+ e^- \tilde\chi_{1}^{0}$ for a speicific value of the dilepton kinematic endpoint.  We performed our analysis in the mSUGRA paradigm with $A_0 = 0$ and $\tan\beta=10$, though this analysis is clearly extendible to more general theories.  We are able to easily determine whether the intermediate slepton is real or virtual.  This provides one clean bound on the slepton mass.  In the case of light virtual sleptons, we find that we can place significant lower and upper bounds on their masses.  For very heavy virtual sleptons, we can only place a lower bound.  However, this bound is generally above $1$ TeV, except for the possibility of cancellation between $Z$ and slepton diagrams with $\mu <0$.

There is still work left to be performed to complete this analysis.  As explained, earlier, we do not expect to find any difficulties in eliminating either the standard model or SUSY background to these distributions, but this is in the process of being verified.  In addition, future analyses will be performed that extend this method to the general MSSM.  Furthermore,  the exact extent to which this measurement assists the determination of the neutralino dark matter density needs to quantified.

\begin{acknowledgments}
A.~B. and K.~.M. wish to acknowledge the kind hospitality of the Aspen Center for Physics during the writing of these proceedings.  The authors were supported by a US DoE OJI award under grant DE-FG02-97ER41029.

\end{acknowledgments}


\end{document}